\documentclass[12pt,onecolumn]{IEEEtran}
\usepackage{amsfonts}
\IEEEoverridecommandlockouts

\ifCLASSINFOpdf
\else
\fi
\usepackage{epstopdf}
\usepackage{multirow}
\usepackage{cite}
\usepackage{stfloats}
\usepackage{color}
\usepackage{epsfig}
\usepackage{graphicx}
\usepackage{psfig}
\usepackage{epsf}
\usepackage{slashbox}
\usepackage{float}
\usepackage{amssymb }
\usepackage[cmex10]{amsmath}
\usepackage{algorithm} 
\usepackage{algorithmic}
\usepackage{booktabs}
\usepackage{amsmath,bm}
\begin{document}
\title{Impacts of Hardware Impairments on Mutualistic Cooperative Ambient Backscatter Communications}
\author{Yinghui Ye, Liqin Shi,    Xiaoli Chu,  Guangyue Lu, and Sumei Sun
\thanks{This work was supported  in part by  the the Natural Science Basic Research Program of Shaanxi under Grant 2021JQ-713,  in part by the Young Talent fund of University Association for Science and Technology in Shaanxi under Grant 20210121, in part by the Scientific Research
Program Funded by Shaanxi Provincial Education Department under Grant 21JK0914.}
\thanks{Yinghui Ye, Liqin Shi and Guangyue  are with the Shaanxi Key Laboratory of
Information Communication Network and Security, Xi'an University of Posts
\& Telecommunications, China. (connectyyh@126.com, liqinshi@hotmail.com, tonylugy@163.com)}
\thanks{Xiaoli Chu is with the Department of Electronic and Electrical Engineering,
The University of Sheffield, U.K. (e-mail: x.chu@sheffield.ac.uk)}
\thanks{Sumei Sun is with the Institute of Infocomm Research, Agency for Science,
Technology and Research, Singapore. (e-mail: sunsm@i2r.a-star.edu.sg)}
}
\markboth{}
{Shi\MakeLowercase{\textit{et al.}}:}
\maketitle
\begin{abstract}
In mutualistic cooperative ambient backscatter communications (AmBC),  Internet-of-Things (IoT) device sends its information to a desired receiver by modulating and backscattering the primary signal, while providing beneficial multipath diversity to the primary receiver   in return, thus forming a mutualism relationship between the AmBC and primary links. We note that the hardware impairments (HIs), which are unavoidable in practical systems and may significantly affect the transmission rates of the primary and AmBC links and  their mutualism relationships, have been largely ignored in the study of mutualistic cooperative AmBC networks. In this paper, we consider a mutualistic cooperative AmBC network under HIs, and study the impacts of HIs on the achievable  rates of the primary link and  the AmBC link. In particular, we theoretically prove that although  HIs degrades the rate of each link,  the mutualism relationship  between the AmBC and  primary links is maintained, i.e., the rate of the primary link in the mutualistic cooperative AmBC network is still higher than that without the AmBC link. The closed-form rate  expressions  of  both the AmBC and primary links are  derived. Computer simulations are provided to validate our theoretical analysis.
\end{abstract}
\IEEEpeerreviewmaketitle
\section{Introduction}
Ambient backscatter communications (AmBC) have been proposed as a spectrum- and energy- efficient solution for Internet of Things (IoT) \cite{8454398,9051982}. The basic principle  of AmBC is to allow an IoT device adjusting the antenna's load impedance to passively modulate  information on the primary  signals   and backscatter the modulated signals to the associated receiver so that the power-hungry components can be avoided in IoT devices \cite{9051982,8730429}. 
AmBC was validated by various practical prototypes \cite{2013Ambient,kellogg2014wi,8103031}. In AmBC, the AmBC receiver receives signals from
both the primary transmitter (PT)  and the AmBC transmitter\footnote{In this paper, the AmBC transmitter and the IoT device are interchangeably used.}  simultaneously, and the PT's signal, e.g., cellular signal, is usually much stronger than that of AmBC. Meanwhile, due to the non-cooperative spectrum sharing between the primary and AmBC link, it is  hard for the AmBC receiver to obtain the channel state information of each link, and hence cannot remove  the severe co-channel interferences caused by the PT's signal \cite{9250656}. Accordingly, the transmission performance of AmBC links is limited and may not meet the requirement of IoT devices.
Cooperative AmBC \cite{9193946}, where the primary link and the AmBC link are jointly designed, has appeared as a promising solution to overcome the above challenge. Depending on whether the AmBC modulation rate  is equal to or much slower than that of PT, cooperative AmBC is classified into  parasitic cooperative AmBC and mutualistic cooperative AmBC. In the former,   the AmBC receiver  firstly decodes the PT's signal while treating the interference from the AmBC link as noise and then removes the decoded PT signal from the composite received signal to decode the AmBC signal \cite{7997001}.
In the latter, as the  AmBC modulation rate is much slower  than that of the PT and the received  AmBC signal includes  both the PT's and  IoT device's information, the AmBC link  provides an additional multipath gain for   the primary receiver to decode  PT's signal,   forming a  mutualism relationship between the primary and AmBC links  \cite{9193946}.

In \cite{8907447}, the authors  maximized the weighted sum rate of both the primary and AmBC links  by  jointly optimizing the PT's transmit power and beamforming vectors in parasitic and mutualistic cooperative AmBC networks, respectively. The similar optimizations were also studied in \cite{9036977} by considering the finite block length in AmBC links. The authors of  \cite{9120210} formulated a stochastic optimization to maximize the  utility function of the signal-to-interference-plus-noise ratio (SINR)  in a parasitic cooperative AmBC network. Considering the energy-causality constraint at each IoT device, the authors in \cite{9461158} jointly optimized the PT's transmit power and the IoT device's power reflection coefficient and  backscattering time to maximize the energy efficiency in parasitic and mutualistic cooperative AmBC networks, respectively. In addition to the above works \cite{8907447, 9036977, 9120210,9461158} with a focus on resource allocation, the performance evaluation  was also investigated  in cooperative AmBC networks. In \cite{9328518}, the outage probability and the diversity gain for a parasitic cooperative AmBC network were derived. Combining parasitic cooperative AmBC with  downlink non-orthogonal multiple
access (NOMA),  the outage probability and ergodic capacity were analyzed theoretically in \cite{8761990}. Considering a mutualistic cooperative AmBC network, the authors in \cite{8807353} derived the upper bounds of the ergodic capacity for  both the primary and AmBC links.

In the above works \cite{8907447, 9036977, 9120210,9461158, 9328518,8761990,8807353}, the radio frequency (RF) front ends of each transceiver are assumed to be ideal, which may be unrealistic.   This is because in practical communications, RF front ends are susceptible to a variety of hardware impairments (HIs), e.g., in-phase/quadrature imbalance, quantization error, etc,  distorting  signals generated by the transmitter  and thus degrading the information decoding  performance at the receiver \cite{9220812}. In spite  of  the efforts on the development of  mitigation algorithms, there always exist residual HIs  due to the time-varying hardware characteristics. Accordingly, HIs should be taken into consideration in the study of cooperative AmBC networks. However, to the best of our knowledge, only in  a very recent work \cite{9319204},  the authors derived the expressions for both the   outage probability  and  intercept probability  in a \emph{parasitic} cooperative NOMA-AmBC network, where the RF front ends of all transceivers suffer HIs. That is, the impacts of HIs on the \emph{mutualistic} cooperative AmBC network  are still unknown. In particular,  when  the HIs exist, the AmBC link not only brings  the additional multipath but also backscatters the distortion noises that are generated by the PT's front ends  to the primary link. More  specifically, if allowing an IoT device to share spectrum with a primary link via AmBC  does not benefit  the primary link's rate, then there is no return for the primary link and  no   mutualism relationship. In this regard, a natural question arises: \emph{Does the  mutualism relationship between the primary and AmBC links still exist in the presence of HIs?}

In this work, we consider a \emph{mutualistic} cooperative AmBC network under HIs, where the IoT device modulates its information on the PT's signal with a much slower modulation rate compared with that of the PT. Our goal is  to reveal the impacts of HIs and answer the above question. The main contributions are summarized as follows.

We theoretically prove the following three results that achieve the above goal. First,  the primary link's rate under HIs is strictly smaller than that under ideal hardware. Second, for given HI parameters,  the primary link's rate that is derived when the AmBC link exists is strictly larger than that without any AmBC link. The above two results validate that the existence of HIs  degrades the rate of both the AmBC and primary links but does not destroy  the  mutualism relationship  between the AmBC and primary links.  Compared with \cite{8907447}, our  conclusion\footnote{The ideal hardware in \cite{8907447} is a special case of our considered model, thus the  conclusion that mutualism relationship  between the AmBC and primary links exists is also valid for the ideal case.} is more rigorous and the details are summarized in Remark 2. Third, we demonstrate  that the HIs lead to the rate ceilings for the primary and AmBC links by deriving the upper-bound rates of the  primary link and that of the  AmBC link at a very high PT's transmit power. Besides, under the assumption that the modulated information of the IoT device follows a symmetric complex Gaussian distribution,   we derive  closed-form rate expressions of both the  primary and AmBC links.

\begin{figure}
  \centering
  \includegraphics[width=0.45\textwidth]{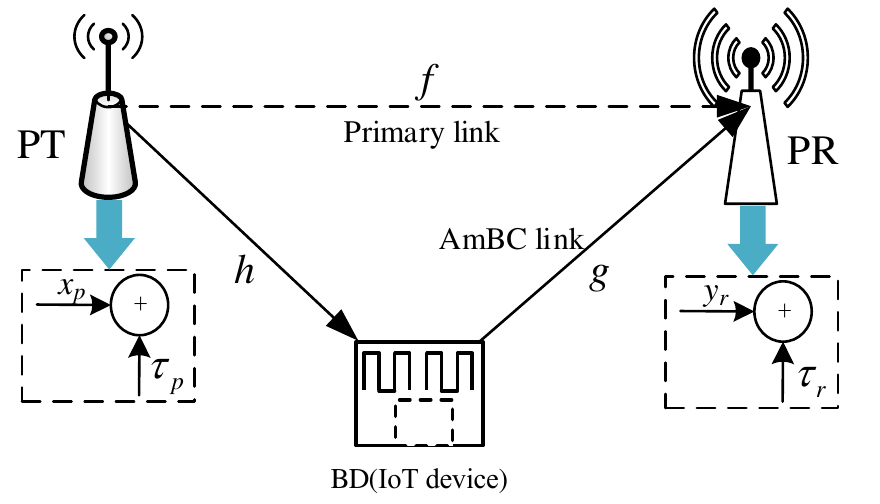}\\
  \caption{System model.}\label{fig0}
\end{figure}
\section{System Model and Rate Analysis}
Fig. 1 is a  mutualistic cooperative AmBC network  that consists of one primary	transmitter (PT), one primary receiver (PR), and one IoT device (also referred to  as a backscatter device (BD) in this paper). Both  PT and  PR are non-energy-constrained   transceivers that are composed of active components, e.g., oscillators. Denote $h$, $f$, $g$ as the channel gains of the PT-BD link, the PT-PR link (also termed as the primary link), and the BD-PR link (also termed as the AmBC link), respectively.  A block-fading channel model is considered, i.e.,  all the channel gains  stay constant within each transmission block but may change across different transmission blocks.   To obtain the performance bound of  both primary and AmBC links, we assume  perfect channel state information.


Let $x_p(n)$ and $c_{s}(i)$ denote the primary signal in the $n$-th symbol period and the AmBC signal in the  $i$-th symbol period, respectively. Due to the simple backscatter circuit in the BD and the low modulation rate of the BD,   the  symbol period of $x_p(n)$, denoted by $T_p$, is  shorter than that of  $c_{s}(i)$, denoted by $T_c$. For analytical tractability, we assume that $T_c=LT_p$, where $L\gg1$ is a positive integer, and  that $x_p(n)$ follows  an independent  circularly symmetric complex Gaussian distribution, i.e., $x_p(n)\sim\mathcal{CN}(0,1)$. Also, the mean and  variance  of $c_{s}(i)$ are assumed to be zero and one, respectively.

In the considered network, PT and BD  work in the cooperative mode and  transmit their information to the PR by sharing  the same resource block. More specifically, PT conveys information to the PR, meanwhile,  BD modulates its own information on the PT's signals and reflects the modulated  signal to the PR. Accordingly, the transmit signal of the PT and the received signal of the BD can be written as, respectively\footnote{Similar to \cite{9250656,9193946}, here we omit the thermal noise at the BD as its power introduced by the passive components is much smaller than that of   $ h\left( {{x_p}(n) + {\tau _p}(n)} \right)$. },
\setcounter{equation}{0}
\begin{align}\label{1}
y_{\rm{PT}}(n)=  x_p(n) + {\tau _p(n)},
\end{align}
\begin{align}\label{2}
y_{\rm{BD}}(n)= \sqrt  h\left( {{x_p}(n) + {\tau _p}(n)} \right),
\end{align}
where ${\tau _p}(n)$ is the distortion noise caused by  HIs of the PT.  $\tau_p(n)$ follows an independent zero-mean  circularly symmetric complex Gaussian distribution,  and its variance is the product of the average power of the PT's signal  $P_0$ and the   HIs level parameter $\kappa_p$ \cite{9220812}, i.e.,  ${\tau _p}(n) \sim \mathcal{CN}\left( {0,\kappa_p^2{P_0}} \right)$.

Accordingly,   the received signal of the PR is expressed as
\begin{align} \label{3}\notag
{y_{{\rm{PR}}}}\left( n \right) &= \underbrace {\sqrt {\beta hg} \left( {{x_p}(n) + {\tau _p}(n)} \right){c_{{s}}}\left(i \right)}_{{\rm{PT}} \to {\rm{BD}} \to {\rm{PR}}\; {\rm{link}}} \\
&+ \underbrace {\sqrt {f}\left( {{x_p}(n) + {\tau _p}(n)} \right)}_{{\rm{PT}} \to {\rm{PR}}\;{\rm{link}}} + {\tau _r}(n) + w\left( n \right),
\end{align}
where  $w\left( n \right)$ is the  additive complex white Gaussian noise with mean zero and variance $\sigma^2$,   $\beta$  is the power reflection coefficient, ${\tau _r}(n) $ is the hardware distortion noise introduced by the PR with the HIs level parameter $\kappa _r$.
The power of ${\tau _r}(n) $, for given ${{\left| {{c_s}(i)} \right|}^2}$,   equals ${\kappa _r^2{P_0}\left( {1 + \kappa _p^2} \right)\left( {h\beta g{{\left| {{c_s}(i)} \right|}^2} + f} \right)}$, while  for each transmission block, the power of ${\tau _r}(n) $ can be calculated as $\kappa _r^2{P_0}\left( {1 + \kappa _p^2} \right)\left( {h\beta g\mathbb{E}\left[ {{{\left| {{c_s}(i)} \right|}^2}} \right] + f} \right)=\kappa _r^2{P_0}\left( {1 + \kappa _p^2} \right)\left( {h\beta g + f} \right)$. Accordingly, the  distribution of ${\tau _r}(n) $ can be written as
\begin{align}\label{5}
{\tau _r}(n)\sim\left\{ {\begin{array}{*{20}{c}}
{{\cal C}{\cal N}\left( {0,\kappa _r^2{P_0}\left( {1 + \kappa _p^2} \right)\left( {h\beta g{{\left| {{c_s}(i)} \right|}^2} + f} \right)} \right),\;{\rm{within}}\;{\rm{one}}\;{\rm{BD}}{\emph{\rq}}{\rm{s}}\;{\rm{symbol}}}\\
{{\cal C}{\cal N}\left( {0,\kappa _r^2{P_0}\left( {1 + \kappa _p^2} \right)\left( {h\beta g + f} \right)} \right),\;{\rm{within}}\;{\rm{one}}\;{\rm{transmission}}\;{\rm{block}}}
\end{array}} \right.
\end{align}

Due to $T_c=LT_p$,  ${c_{{s}}}\left(i \right)$ spans $L$ primary symbol periods for $n=1,2,...,L$, i.e., ${c_{{s}}}\left(i \right)$  keeps almost unchanged for decoding ${x_p}(n)$, $n=1,2,...,L$. For a given ${c_{{s}}}\left(i \right)$, the first  term in \eqref{3} can be rewritten as $\sqrt {\beta hg} {x_p}(n){c_{{s}}}(i) + \sqrt {\beta hg} {\tau _p}(n){c_{{s}}}(i)$, where $\sqrt {\beta hg} {x_p}(n){c_{{s}}}(i)$ and $\sqrt {\beta hg} {\tau _p}(n){c_{{s}}}(i)$ can be regarded as the output
of the PT signal ${x_p}(n)$ passing through a slowly varying channel $\sqrt {\beta hg} {c_{{s}}}(i)$ and the Gaussian noise with variance $\kappa_p^2{P_0}{\beta hg}|{c_{{s}}}(i)|^2$, respectively. Accordingly,  the SINR to decode ${x_p}(n)$ and the achievable rate of ${x_p}(n)$ within the  symbol period of ${c_{{s}}}\left(i \right)$ can be calculated as, respectively,
\setcounter{equation}{4}
\begin{align}\label{6a}
{\gamma _{{x_p}}}\left( {{c_s}(i)} \right)={\frac{{{P_0}h\beta g{{\left| {{c_s}(i)} \right|}^2} + {P_0}f}}{{{P_0}\left( {h\beta g{{\left| {{c_s}(i)} \right|}^2} + f} \right)\kappa  + {\sigma ^2}}}},
\end{align}
\begin{align}\label{6b}
{R_p}\left( {{c_s}(i)} \right) = B_w{\log _2}\left( 1+{\gamma _{{x_p}}}\left( {{c_s}(i)} \right) \right),
\end{align}
where  ${\kappa {\rm{ = }}\kappa _r^2\kappa _p^2 + \kappa _r^2 + \kappa _p^2}$, and $B_w$ denotes the communication bandwidth.

Assuming that the number of symbols of the BD signal  is sufficiently large within one transmission block, then the average rate of ${x_p}(n)$  is given as
\begin{align}\label{7}
\mathcal{{C}}_p={\mathbb{E}_{{c_{{s}}}(i)}}\left[ {R_p\left( {{c_{{s}}}(i)} \right)} \right],
\end{align}
where ${\mathbb{E}_x}[ \cdot ]$ denotes the expectation operator  over the random variable $x$.

After obtaining the rate of the primary link, we put our attention on the rate  of the AmBC link.  As one of main focuses in this work is to see whether the existence of HIs destroys the mutualism transmission in the mutualistic cooperative AmBC network or not, we assume for simplicity that   $\sqrt {hf} {x_p}(n)$ can be perfectly removed from ${y_{{\rm{PR}}}}\left( n \right)$ via successive interference cancellation (SIC). Thus, the remaining signal at the PR to decode ${c_{{s}}}(i)$ is given by
\begin{align}\label{8}\notag
{\hat y_{{\rm{PR}}}}\left( n \right) &= \sqrt {\beta hg} \left( {{x_p}(n) + {\tau _p}(n)} \right){c_{s}}\left( i \right)\\
 &+ \sqrt {f} {\tau _p}(n) + {\tau _r}(n) + w\left( n \right).
\end{align}

As the average power of ${x_p}(n)$ equals one and  one BD symbol ${c_{s}}\left( i \right)$ is modulated into the $L$ consecutive PT symbol periods ${x_p}(n)$, the average SINR to decode ${c_{s}}(i)$ via maximal ratio combing (MRC) of ${\hat y_{{\rm{PR}}}}\left( n \right)$, $n=1,2,...,L$, which are received in $L$ consecutive primary symbol periods,  in each transmission block, can be approximated as
\begin{align} \label{9} 
{\gamma _{{c_{s}}}} =  \sum\limits_{n = 1}^L {\mathbb{E}\left[ {\frac{{\beta hg{{\left| {{x_p}(n)} \right|}^2}}}{{{{\left| {{\tau _p}(n)} \right|}^2}\left( {\beta hg + f} \right) + {{\left| {{\tau _r}(n)} \right|}^2} + {{\left| {w\left( n \right)} \right|}^2}}}} \right]}= \frac{{L\beta hg{P_0}}}{{{P_0}\left( {\beta hg + f} \right)\kappa  + {\sigma ^2}}}.
\end{align}
In the mutualistic cooperative AmBC network, as modulating one BD symbol requires
$L$ consecutive PT symbols, the PT's signal
$x_p(n)$ can be viewed as a spread-spectrum code with length
$L$ for BD symbols. Accordingly, the SINR to decode ${c_{s}}(i)$ is increased by $L$ times  at the price of symbol rate decreased by $\frac{1}{L}$, and the BD's rate can be expressed as
\begin{align} \label{10}
{\mathcal{{C}}_s} = \frac{B_w}{L}{\log _2}\left( {1 + {\gamma _{{c_{s}}}}} \right).
\end{align}
\section{Impacts of HIs on Rate}
The  rates for the primary link and the AmBC link with ideal  hardware can be obtained by substituting $\kappa _p=\kappa _r=0$ into \eqref{7} and \eqref{10}, respectively, given as
\begin{align}\label{11}
\mathcal{{C}}_p^{{\rm{id}}} =\mathbb{E }{_{{c_s}(i)}}\left[B_w {{{\log }_2}\left( {1 + \frac{{{P_0}h\beta g{{\left| {{c_s}(i)} \right|}^2} + {P_0}f}}{{{\sigma ^2}}}} \right)} \right],
\end{align}
\begin{align}\label{12}
\mathcal{{C}} _{s}^{{\rm{id}}} = \frac{B_w}{L}{\log _2}\left( {1 + \frac{{L{P_0}\beta hg}}{{{\sigma ^2}}}} \right).
\end{align}

Comparing the rates in \eqref{11} and \eqref{12}  with the HIs case in \eqref{7} and \eqref{10}, it is clear that  the  rates with HIs are strictly  smaller than those of ideal hardware, i.e., $\mathcal{{C}}_p^{{\rm{id}}}>\mathcal{{C}}_p$ and ${\cal{R}} _{s}^{{\rm{id}}} >{\cal{R}} _{s} $ always hold when $\kappa>0$. This indicates that the existence of HIs  degrades the achievable rates of both the  primary and AmBC links. Besides, by assuming ${P_0} \to \infty $ in the case of HIs, we have the following inequality associated with the PT's rate, i.e.,
\begin{align}\label{13} \notag
\mathcal{{C}}_p&\leq{B_w\log _2}\left( {1 + \frac{{h\beta g\mathbb{E}\left[ {{{\left| {{c_s}(i)} \right|}^2}} \right] + f}}{{\left( {h\beta g\mathbb{E}\left[ {{{\left| {{c_s}(i)} \right|}^2}} \right] + f} \right)\kappa + {\frac{{{\sigma ^2}}}{{{P_0}}}}}}} \right) 
\\&<B_w{\log _2}\left( {1 + \frac{1}\kappa} \right),
\end{align}
where the first inequality holds for the Jensen's inequality, and  second inequality is derived from that ${ \frac{{h\beta g\mathbb{E}\left[ {{{\left| {{c_s}(i)} \right|}^2}} \right] + f}}{{\left( {h\beta g\mathbb{E}\left[ {{{\left| {{c_s}(i)} \right|}^2}} \right] + f} \right)\kappa + {\frac{{{\sigma ^2}}}{{{P_0}}}}}}}$ is an increasing function with respect to $P_0$ and that ${\frac{{{\sigma ^2}}}{{{P_0}}}}$ approaches to zero as ${P_0} \to \infty $. 

Similar as above, we can obtain the following inequality on the BD's rate as ${P_0} \to \infty $, given by
\begin{align}\label{14}
\mathcal{{C}}_s&<\frac{B_w}{L}{\log _2}\left( {1 + \frac{{L\beta hg}}{{\left( {\beta hg + f} \right)\kappa}}} \right)
\end{align}

{\emph{Remark 1.}} Both   \eqref{13} and \eqref{14} show that the HIs level parameter have significant impacts on the achievable rate. It can be seen that in the case of HIs,
  the achievable rates of both the  primary and AmBC links  are bounded at ${P_0} \to \infty $, i.e., there exist rate ceilings for both the  primary and AmBC links. Interestingly, the upper bound of  $\mathcal{{C}}_p$ {\emph{only}} depends on the HIs level parameter, while for the AmBC link, its upper bound is affected by  the HIs level parameter and the channel gains. Particularly, the upper bound of $\mathcal{{C}}_s$ increases with the decrease of $f$, indicating that a poor channel condition of the PT-PR link raises the ceiling of the AmBC link rate.


 We note that in this seminal contribution \cite{8907447}, it has been shown that  in the ideal hardware case, allowing BD to share the same spectrum with the PT can   offer multipath diversity to the primary link without introducing any harmful factors and thus  the primary link achieves a larger  transmission rate than the case where the spectrum resource is alone  used by the PT,  yielding the mutualism transmission between primary and AmBC links. However, in the presence of HIs, the AmBC link not only provides  beneficial  multipath to the primary link, but also brings the extra hardware distortion noise that is a harmful factor to the primary link's rate. In this regard, a natural question arises: does the mutual benefit still hold  in the presence of HIs? To answer this question, we provide the following theorem.

{\bf{Theorem 1.}} For given  $\kappa$ and all channel gains, we have the following inequality, i.e.,
\begin{align}\label{15}
{\mathcal{{C}}_p} >B_w {\log _2}\left( {1 + \frac{{{P_0}f}}{{{P_0}f\kappa + {\sigma ^2}}}} \right),
\end{align}
where the right side of \eqref{15} is the achievable rate when the spectrum resource is alone  used by the PT, i.e., access denied for BD, in the case of  HIs.

\emph{Proof.} Please refer to Appendix A.   \hfill {$\blacksquare $}

\emph{Remark 2.} Theorem 1 indicates that in the mutualistic cooperative AmBC network with HIs, even though the access of BDs brings both multipath diversity and   hardware distortion noise to the primary link,  the rate of the primary link can be still improved. That is, the existence of HIs  does not destroy the  mutual benefit between the AmBC and primary links. In particular, letting   $\kappa_p=\kappa _r=0$ ($\kappa=\kappa _r^2\kappa _p^2 + \kappa _r^2 + \kappa _p^2=0$),  Theorem 1 also verifies that the mutual benefit exists in  the ideal hardware case. We note that the existence of  mutual benefit in the ideal hardware case was also proven in \cite{8907447}, where the assumption that ${{c_s}(i)}$ follows the complex Gaussian distribution and the approximation under high signal-noise-ratio (SNR) were adopted, however,  in our work, we have not made any special assumptions on the distribution of ${{c_s}(i)}$  and also have not used any approximations. Thus, our proposed Theorem 1 is more general and rigorous compared to  the existing one \cite{8907447}.

Although we provide expressions to calculate ${\mathcal{{C}}_p}$ and its upper bound, \eqref{7} is not in the closed form and   \eqref{13} is tight only at ${P_0} \to \infty $. This indicates that  the upper bound in \eqref{13} does not hold at low or moderate  transmit power. Accordingly, it is required to derive a closed-form expression  that  approximates ${\mathcal{{C}}_p}$ well no matter what the transmit power is. To this end,  Proposition 1 is provided.

\textbf{Proposition 1.} By assuming  that ${{c_s}(i)}$  follows the symmetric complex Gaussian distribution with  zero mean and unit variance, we can approximate ${\mathcal{{C}}_p}$ as
\begin{align}\label{16}
{\mathcal{{C}}_p}=\left\{ {\begin{array}{*{20}{c}}
\!\!\!\!\!\!\!\!\!\!\!\!\begin{array}{l}
{B_w}{\log _2}\left( {\frac{{b + b\kappa  + {\sigma ^2}}}{{b\kappa  + {\sigma ^2}}}} \right) - \frac{{{B_w}}}{{\ln 2}}\exp \left( {\frac{{b + b\kappa  + {\sigma ^2}}}{{a\kappa {\rm{ + }}a}}} \right){\rm{Ei}}\left( { - \frac{{b + b\kappa  + {\sigma ^2}}}{{a\kappa {\rm{ + }}a}}} \right)\\
 + \frac{{{B_w}}}{{\ln 2}}\exp \left( {\frac{{b\kappa  + {\sigma ^2}}}{{a\kappa }}} \right){\rm{Ei}}\left( { - \frac{{b\kappa  + {\sigma ^2}}}{{a\kappa }}} \right),{\rm{if}}\;\kappa  > 0
\end{array}\\
{{B_w}{{\log }_2}\left( {\frac{{b + {\sigma ^2}}}{{{\sigma ^2}}}} \right) - \frac{{{B_w}}}{{\ln 2}}\exp \left( {\frac{{b + {\sigma ^2}}}{a}} \right){\rm{Ei}}\left( { - \frac{{b + {\sigma ^2}}}{a}} \right),\;\;{\rm{if}}\;\kappa  = 0\;}
\end{array}} \right.,
\end{align}
 where  $a = {P_0}h\beta g$, $b={P_0}f$, and ${\rm{Ei}}\left(  \cdot  \right)$ is the exponential integral.

\emph{Proof.} Please refer to Appendix B. \hfill {$\blacksquare $}

\emph{Remark 3.} By comparisons between \eqref{16} and the right side of \eqref{15}, one can see that the increased rates of the primary link in the case of HIs and ideal hardware case are $\frac{B_w}{{\ln 2}}\exp \left( {\frac{{b\kappa  + {\sigma ^2}}}{{a\kappa }}} \right){\rm{Ei}}\left( { - \frac{{b\kappa  + {\sigma ^2}}}{{a\kappa }}} \right) - \frac{B_w}{{\ln 2}}\exp \left( {\frac{{b + b\kappa  + {\sigma ^2}}}{{a\kappa {\rm{ + }}a}}} \right){\rm{Ei}}\left( { - \frac{{b + b\kappa  + {\sigma ^2}}}{{a\kappa {\rm{ + }}a}}} \right)$ and $ - \frac{B_w}{{\ln 2}}\exp \left( {\frac{{b + {\sigma ^2}}}{a}} \right) {\rm{Ei}}\left( { - \frac{{b + {\sigma ^2}}}{a}} \right)$, respectively. It can  be inferred that the HIs degrade the increased rate, as $ - \frac{B_w}{{\ln 2}}\exp \left( {\frac{{b + {\sigma ^2}}}{a}} \right){\rm{Ei}}\left( { - \frac{{b + {\sigma ^2}}}{a}} \right)>\frac{B_w}{{\ln 2}}\exp \left( {\frac{{b\kappa  + {\sigma ^2}}}{{a\kappa }}} \right){\rm{Ei}}\left( { - \frac{{b\kappa  + {\sigma ^2}}}{{a\kappa }}} \right)- \frac{B_w}{{\ln 2}}{\rm{Ei}}\left( { - \frac{{b + b\kappa  + {\sigma ^2}}}{{a\kappa {\rm{ + }}a}}} \right) \exp \left( {\frac{{b + b\kappa  + {\sigma ^2}}}{{a\kappa {\rm{ + }}a}}} \right)$ holds at $\kappa>0$. This indicates that although the existence of HIs does not destroy the mutual benefit between the primary and AmBC links, the harmful impacts on the increased rate  exist.
\section{Simulations}
In this section, computer simulations are provided to support our findings.
Unless otherwise specified, the basic simulation parameters are set as follows.
In particular, we set $P_0=3$ mW, $T=1$ s, $B_w=1$ MHz,  $L=128$ ,   $\kappa_p=\kappa_r=0.1$, and the noise power spectral density is set as $\sigma^2=-120$ dBm/Hz \cite{8907447,9319204}. The standard channel fading model is considered, where each channel gain is given by the product of the small-scale fading and the large-scale fading.
Let $D_{\rm{ps}}$, $D_{\rm{sr}}$ and $D_{\rm{pr}}$ denote the distances of the PT-BD link, the BD-PR link and the PT-PR link, respectively.
Denote $h'$, $f'$ and $g'$ as the small-scale fading of the PT-BD link,  the BD-PR link and the PT-PR link, respectively.
Accordingly, we have $h=h'D_{\rm{ps}}^{-\alpha_{\rm{ps}}}$, $f=f'D_{\rm{sr}}^{-\alpha_{\rm{sr}}}$ and $g=g'D_{\rm{pr}}^{-\alpha_{\rm{pr}}}$, where $\alpha_{\rm{ps}}$, $\alpha_{\rm{sr}}$ and $\alpha_{\rm{pr}}$ are the path loss exponents of the PT-BD link,   the BD-PR link and the PT-PR link, respectively.
Here $\alpha_{\rm{ps}}$, $\alpha_{\rm{sr}}$ and $\alpha_{\rm{pr}}$ are set as $2.7$, $2.7$ and $3$, respectively.
The BD's power reflection coefficient $\beta$ is set as $0.8$.

\begin{figure}
  \centering
  \includegraphics[width=0.4\textwidth]{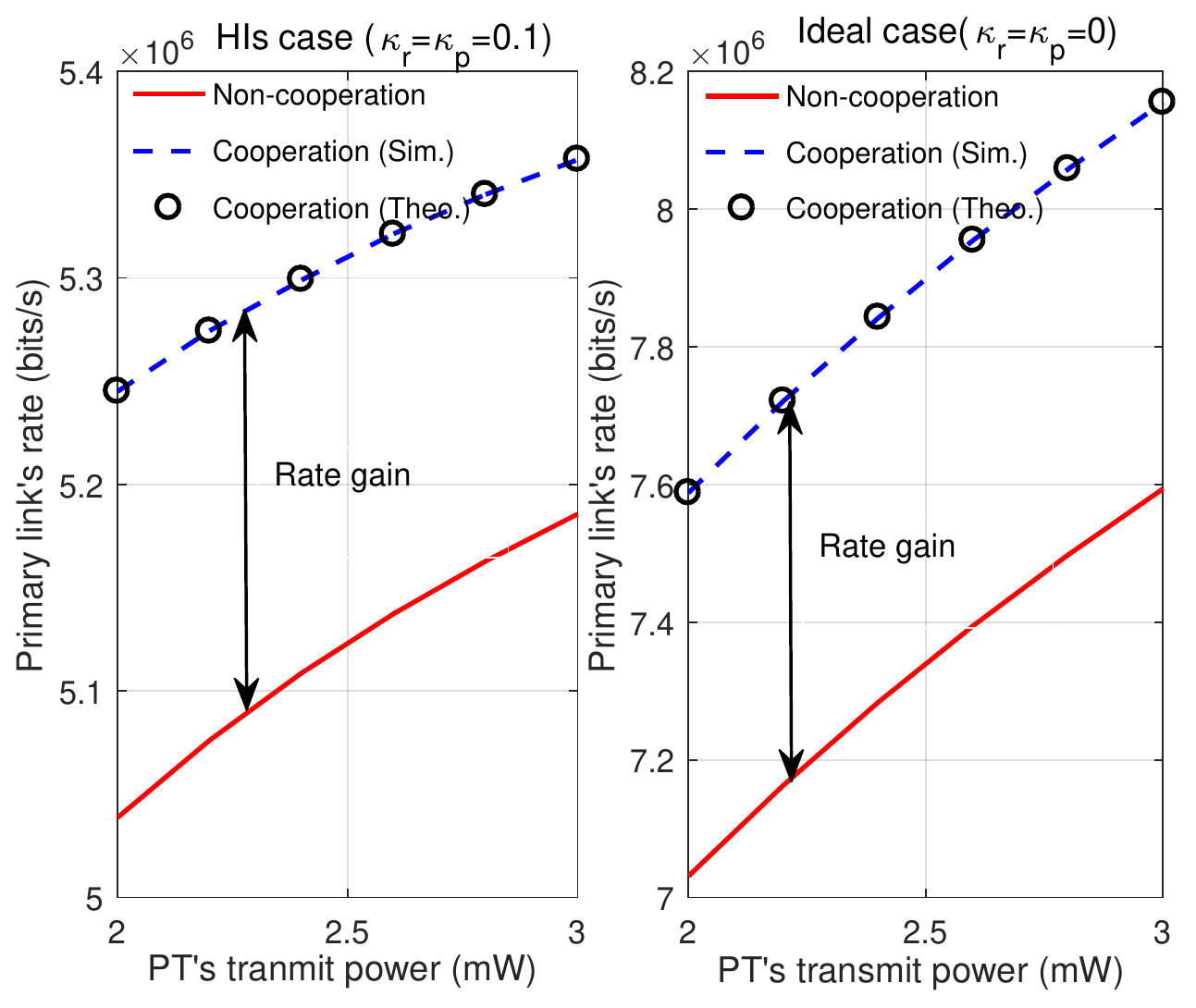}\\
  \caption{The PT's rate versus the transmit power of the PT $P_0$ under the HIs case and the ideal case.}\label{fig3}
\end{figure}

Fig. 2 shows the PT's rate versus the transmit power of the PT under the HIs case and the ideal case.
In order to illustrate the improvement of the PT's rate caused by the BD's cooperation, we compare the PT's transmit rate under the considered network (called \lq\lq Cooperation" in this figure)
with the PT's rate under the network without the BD (called \lq\lq Non-cooperation" in this figure).
For the PT's rate under the considered network, we plot the simulation results and the theoretical results, respectively, where
the simulation results are obtained via Monte Carlo simulations (marked by \lq o') averaged over $1\times10^6$  realizations
and the theoretical results are achieved based on the derived expression \eqref{16}.
It can be observed that the theoretical results match well with the simulation results, which demonstrates the correctness of \eqref{16}.
By comparisons, we can see that the PT's rate under the considered network is always higher than that without the BD's cooperation in both the HIs case and the ideal case, which
verifies Theorem 1, indicating that although the HIs level has significant impacts on the PT's rate, the improvement from the BD's cooperation still exists.
Besides, by comparing the rate gains under the HIs case and the ideal case, we can also find that the existence of the HIs not only reduces the achievable rate at the PT, but also degrades the rate gain.

Fig. 3 illustrates the impacts of the HIs level parameters on the PT's rate. Here we set the values of $\kappa_r$ and $\kappa_p$ are same and vary from $0$ to $0.2$.
It can be observed that the simulation results with the BD's cooperation (marked by \lq o') always match well with the theoretical results via \eqref{16},
indicating the correctness of theoretical derivations.
We can also see that the PT's rate with/without the BD's cooperation decrease when the HIs level parameters increase, since the existence of the HIs
degrades the achievable rate at the PT and the larger the HIs level parameters are, the smaller the PT's rate is.
Besides, the rate gain from BD's cooperation always exists no matter what the HIs level is and a larger HIs level brings a smaller rate gain.

\begin{figure}
  \centering
  \includegraphics[width=0.425\textwidth]{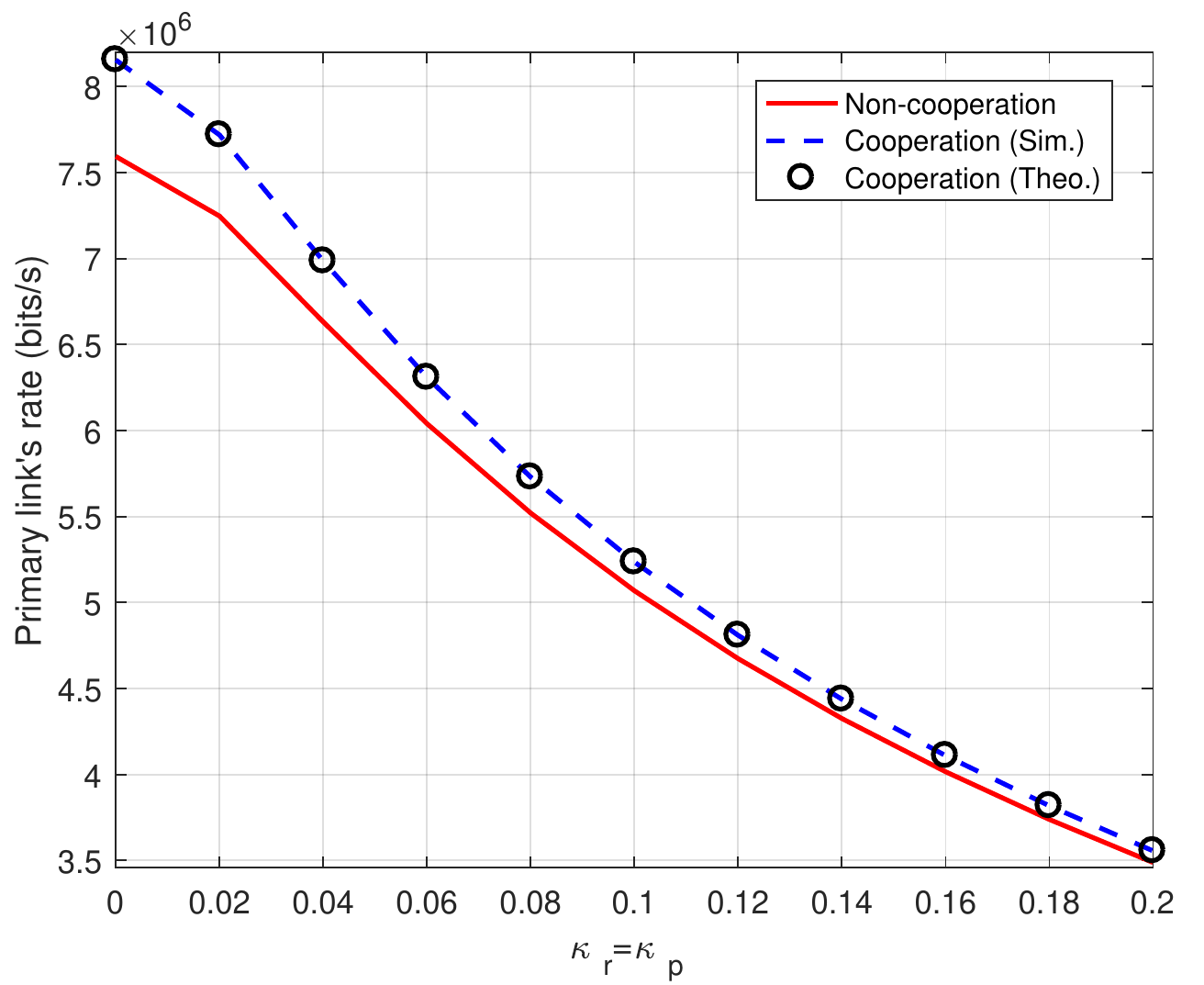}\\
  \caption{The impacts of the HIs level parameters on the PT's rate.}\label{fig4}
\end{figure}
\section{Conclusions}
In this paper, we have investigated the impacts of HIs on the  mutualistic cooperative AmBC network. We have proven that the rates of both primary and AmBC links decrease in the presence of HIs and that there exist rate ceilings for both the   primary and AmBC links in the presence of  HIs. We have  validated that the HIs do not destroy the mutualism relationship between the AmBC and primary link and have also derived closed-form rate expressions for both the  primary and AmBC links under the assumption that the BD's message follows a  symmetric complex Gaussian
distribution.  Simulation results have validated our derived results.
\section*{Appendix A}
Let define the following function, i.e., $\phi \left( x \right) = \frac{{{P_0}h\beta gx + {P_0}f}}{{{P_0}\left( {h\beta gx + f} \right)\kappa^2 + {\sigma ^2}}}$. Taking the   first derivative of $\phi \left( x \right)$ with respect to $x$, we have
\begin{align} \label{a1}
\phi '\left( x \right){\rm{ = }}\frac{{P_0^2h\beta g{\sigma ^2}}}{{{{\left( {{P_0}\left( {h\beta gx + f} \right)\kappa^2 + {\sigma ^2}} \right)}^2}}} \tag{A.1}
\end{align}
It is clear from \eqref{a1} that  $\phi \left( x \right)$ increases with $x$ at $x\geq 0$. Thus, we have
\begin{align}\label{a2}
\left\{ {\begin{array}{*{20}{c}}
\begin{array}{l}
{\log _2}\left( {1 + \frac{{{P_0}h\beta g{{\left| {{c_s}(i)} \right|}^2} + {P_0}f}}{{{P_0}\left( {h\beta g{{\left| {{c_s}(i)} \right|}^2} + f} \right)\kappa^2 + {\sigma ^2}}}} \right)\\
 > {\log _2}\left( {1 + \frac{{{P_0}f}}{{{P_0}f\kappa^2 + {\sigma ^2}}}} \right),\:{\rm{if}}\:{\left| {{c_s}(i)} \right|^2} > 0,
\end{array}\\
\begin{array}{l}
{\log _2}\left( {1 + \frac{{{P_0}h\beta g{{\left| {{c_s}(i)} \right|}^2} + {P_0}f}}{{{P_0}\left( {h\beta g{{\left| {{c_s}(i)} \right|}^2} + f} \right)\kappa^2 + {\sigma ^2}}}} \right)\\
 = {\log _2}\left( {1 + \frac{{{P_0}f}}{{{P_0}f\kappa^2 + {\sigma ^2}}}} \right),\:{\rm{if}}\:{\left| {{c_s}(i)} \right|^2} = 0.
\end{array}
\end{array}} \right.\tag{A.2}
\end{align}
In what follows, we examine the value of ${{\left| {{c_s}(i)} \right|}^2}$.
In modulation schemes, different values of  ${{c_s}(i)}$ are mapped to different information. If ${{\left| {{c_s}(i)} \right|}^2}=0$ holds for all $i$, BD cannot modulate its information to the PT's signal and the achievable rate of BD  equals zero. Therefore, in order to achieve the information transmission of the AmBC link,  ${{\left| {{c_s}(i)} \right|}^2}=0$ cannot be always satisfied for any given $i$. Combining this fact,  ${{\left| {{c_s}(i)} \right|}^2}\geq0$ and \eqref{a2}, we reach the following result, given as
\begin{align}\notag \label{a3}
{\mathcal{{C}}_p}&=\mathbb{E}{_{{c_s}(i)}}\left[ B_w{{{\log }_2}\left( {1 + \frac{{{P_0}h\beta g{{\left| {{c_s}(i)} \right|}^2} + {P_0}f}}{{{P_0}\left( {h\beta g{{\left| {{c_s}(i)} \right|}^2} + f} \right)\kappa^2 + {\sigma ^2}}}} \right)} \right]\\ \notag
&  >  {\mathbb{E}_{{c_s}(i)}}\left[B_w {{{\log }_2}\left( {1 + \frac{{{P_0}f}}{{{P_0}f\kappa^2 + {\sigma ^2}}}} \right)} \right]\\
&= B_w{\log _2}\left( {1 + \frac{{{P_0}f}}{{{P_0}f\kappa^2 + {\sigma ^2}}}} \right).\tag{A.3}
\end{align}

Theorem 1 can be proven by \eqref{a3} and the proof is complete.
\section*{Appendix B}
As ${{c_s}(i)}$ obeys a standard symmetric complex Gaussian distribution, the distribution of ${{\left| {{c_s}(i)} \right|}^2}$ is an exponential function with parameter one. Based on this, \eqref{7} can be rewritten as
\begin{align}\notag
\mathcal{{C}}_p&={\mathbb{E}_{{c_{{s}}}(i)}}\left[ {R_p\left( {{c_{{s}}}(i)} \right)} \right]\\ \notag
&\mathop {\rm{ = }}\limits^{x = {{\left| {{c_s}(i)} \right|}^2}} \int_0^\infty  B_w{{{\log }_2}\left( {1 + \frac{{ax + b}}{{a\kappa x + b\kappa  + {\sigma ^2}}}} \right)} \exp \left( { - x} \right)dx\\ \notag
&=\underbrace {\int_0^\infty  {B_w{{\log }_2}\left( {\left( {a\kappa {\rm{ + }}a} \right)x + b + b\kappa  + {\sigma ^2}} \right)} \exp \left( { - x} \right)dx}_{{\Delta _1}} \\
&- \underbrace {\int_0^\infty  {B_w{{\log }_2}\left( {{a\kappa }x + b\kappa  + {\sigma ^2}} \right)} \exp \left( { - x} \right)dx}_{{\Delta _2}}, \tag{B.1}
\end{align}
where $a = {P_0}h\beta g$ and $b={P_0}f$.
Using integration by parts, we have
\begin{align}\notag
{\Delta _1} &= \left. { - B_w{{\log }_2}\left( {b + b\kappa  + {\sigma ^2} + \left( {a\kappa {\rm{ + }}a} \right)x} \right)\exp \left( { - x} \right)} \right|_0^\infty \\ \notag
 &+ \frac{{B_w\left( {a\kappa {\rm{ + }}a} \right)}}{{\ln 2}}\int_0^\infty  {\frac{{\exp \left( { - x} \right)}}{{b + b\kappa  + {\sigma ^2} + \left( {a\kappa {\rm{ + }}a} \right)x}}dx} \\ \notag
& =B_w {\log _2}\left( {b + b\kappa  + {\sigma ^2}} \right) + \frac{B_w}{{\ln 2}}\int_0^\infty  {\frac{{{e^{ - x}}}}{{\frac{{b + b\kappa  + {\sigma ^2}}}{{a\kappa {\rm{ + }}a}} + x}}dx} \\  \notag
&=B_w{\log _2}\left( {b + b\kappa  + {\sigma ^2}} \right) - \frac{B_w}{{\ln 2}}\exp \left( {\frac{{b + b\kappa  + {\sigma ^2}}}{{a\kappa {\rm{ + }}a}}} \right)\\
&\times{\rm{Ei}}\left( { - \frac{{b + b\kappa  + {\sigma ^2}}}{{a\kappa {\rm{ + }}a}}} \right),\tag{B.2}
\end{align}
where the last equality is derived from $\int_0^\infty  {\frac{{\exp \left( { - \mu x} \right)dx}}{{x + \beta }}} dx =  - \exp \left( {\mu \beta } \right){\rm{Ei}}\left( { - \mu \beta } \right)$, as shown in eq.(3.352.4) of  \cite{b1}.

Similar as above, ${\Delta _2}$ can be calculated as 
\begin{align}\notag
{\Delta _2}&=\left. { - B_w{{\log }_2}\left( {a\kappa x + b\kappa  + {\sigma ^2}} \right){e^{ - x}}} \right|_0^\infty  + \frac{B_w}{{\ln 2}}\int_0^\infty  {\frac{{{e^{ - x}}}}{{a\kappa x + b\kappa  + {\sigma ^2}}}} dx\\  \notag
&=\left\{ {\begin{array}{*{20}{c}}
{B_w{{\log }_2}\left( {b\kappa  + {\sigma ^2}} \right) + \frac{B_w}{{a\kappa \ln 2}}\int_0^\infty  {\frac{{{e^{ - x}}}}{{x + \frac{{b\kappa  + {\sigma ^2}}}{{a\kappa }}}}} dx,\;{\rm{if}}\;\kappa  > 0\;}\\
{B_w{{\log }_2}\left( {b\kappa  + {\sigma ^2}} \right),\;\;\;\;\;\;\;\;\;\;\;\;\;\;\;\;\;\;\;\;\;\;\;\;\;\;\;\;\;\;\;\;\;\;\;\;{\rm{if}}\;\kappa  = 0}
\end{array}} \right.\\
&={B_w\log _2}\left( {b\kappa  + {\sigma ^2}} \right) - \left\{ {\begin{array}{*{20}{c}}\!\!\!
{\frac{B_w}{{\ln 2}}\exp \left( {\frac{{b\kappa  + {\sigma ^2}}}{{a\kappa }}} \right){\rm{Ei}}\left( { - \frac{{b\kappa  + {\sigma ^2}}}{{a\kappa }}} \right),\;{\rm{if}}\;\kappa  > 0\;}\\
\!\!\!{0,\;\;\;\;\;\;\;\;\;\;\;\;\;\;\;\;\;\;\;\;\;\;\;\;\;\;\;\;\;\;\;\;\;\;\;\;\;\;\;\;\;\;\;{\rm{if}}\;\kappa  = 0}
\end{array}} \right..\tag{B.3}
\end{align}
Substituting (B.2) and  (B.3) into (B.1), we can reach \eqref{16} and  the proof is complete.

\ifCLASSOPTIONcaptionsoff
  \newpage
\fi
\bibliographystyle{IEEEtran}
\bibliography{refa}

\begin{thebibliography}{10}
\providecommand{\url}[1]{#1}
\csname url@samestyle\endcsname
\providecommand{\newblock}{\relax}
\providecommand{\bibinfo}[2]{#2}
\providecommand{\BIBentrySTDinterwordspacing}{\spaceskip=0pt\relax}
\providecommand{\BIBentryALTinterwordstretchfactor}{4}
\providecommand{\BIBentryALTinterwordspacing}{\spaceskip=\fontdimen2\font plus
\BIBentryALTinterwordstretchfactor\fontdimen3\font minus
  \fontdimen4\font\relax}
\providecommand{\BIBforeignlanguage}[2]{{%
\expandafter\ifx\csname l@#1\endcsname\relax
\typeout{** WARNING: IEEEtran.bst: No hyphenation pattern has been}%
\typeout{** loaded for the language `#1'. Using the pattern for}%
\typeout{** the default language instead.}%
\else
\language=\csname l@#1\endcsname
\fi
#2}}
\providecommand{\BIBdecl}{\relax}
\BIBdecl

\bibitem{8454398}
C.~Xu, L.~Yang, and P.~Zhang, ``Practical backscatter communication systems for
  battery-free internet of things: A tutorial and survey of recent research,''
  \emph{IEEE Signal Process. Mag.}, vol.~35, no.~5, pp. 16--27, 2018.

\bibitem{9051982}
Y.~Ye, L.~Shi, X.~Chu, and G.~Lu, ``On the outage performance of ambient
  backscatter communications,'' \emph{IEEE Internet Things J.}, vol.~7, no.~8,
  pp. 7265--7278, 2020.

\bibitem{8730429}
Y.~Ye, L.~Shi, R.~Qingyang~Hu, and G.~Lu, ``Energy-efficient resource
  allocation for wirelessly powered backscatter communications,'' \emph{IEEE
  Commun. Lett.}, vol.~23, no.~8, pp. 1418--1422, 2019.

\bibitem{2013Ambient}
V.~{Liu} \emph{et~al.}, ``Ambient backscatter: wireless communication out of
  thin air,'' in \emph{Proc. ACM SIGCOMM}, 2013, p. 12–16.

\bibitem{kellogg2014wi}
B.~Kellogg \emph{et~al.}, ``{Wi-Fi} backscatter: Internet connectivity for
  {RF}-powered devices,'' in \emph{Proc. ACM SIGCOMM}, 2014, pp. 607--618.

\bibitem{8103031}
S.~N. Daskalakis, J.~Kimionis, A.~Collado, G.~Goussetis, M.~M. Tentzeris, and
  A.~Georgiadis, ``Ambient backscatterers using fm broadcasting for low cost
  and low power wireless applications,'' \emph{IEEE Trans. Microw. Theory
  Techn.}, vol.~65, no.~12, pp. 5251--5262, 2017.

\bibitem{9250656}
C.~Liu, Z.~Wei, D.~W.~K. Ng, J.~Yuan, and Y.-C. Liang, ``Deep transfer learning
  for signal detection in ambient backscatter communications,'' \emph{IEEE
  Trans. Wireless Commun.}, vol.~20, no.~3, pp. 1624--1638, 2021.

\bibitem{9193946}
Y.-C. Liang, Q.~Zhang, E.~G. Larsson, and G.~Y. Li, ``Symbiotic radio:
  Cognitive backscattering communications for future wireless networks,''
  \emph{IEEE Trans. Cogn. Commun. Netw.}, vol.~6, no.~4, pp. 1242--1255, 2020.

\bibitem{7997001}
X.~Kang, Y.-C. Liang, and J.~Yang, ``Riding on the primary: A new spectrum
  sharing paradigm for wireless-powered {IoT} devices,'' in \emph{Proc. IEEE
  ICC}, 2017, pp. 1--6.

\bibitem{8907447}
R.~Long, Y.-C. Liang, H.~Guo, G.~Yang, and R.~Zhang, ``Symbiotic radio: A new
  communication paradigm for passive internet of things,'' \emph{IEEE Internet
  Things J.}, vol.~7, no.~2, pp. 1350--1363, 2020.

\bibitem{9036977}
Z.~Chu, W.~Hao, P.~Xiao, M.~Khalily, and R.~Tafazolli, ``Resource allocations
  for symbiotic radio with finite blocklength backscatter link,'' \emph{IEEE
  Internet Things J.}, vol.~7, no.~9, pp. 8192--8207, 2020.

\bibitem{9120210}
X.~Chen, H.~V. Cheng, K.~Shen, A.~Liu, and M.-J. Zhao, ``Stochastic transceiver
  optimization in multi-tags symbiotic radio systems,'' \emph{IEEE Internet
  Things J.}, vol.~7, no.~9, pp. 9144--9157, 2020.

\bibitem{9461158}
H.~Yang, Y.~Ye, K.~Liang, and X.~Chu, ``Energy efficiency maximization for
  symbiotic radio networks with multiple backscatter devices,'' \emph{IEEE Open
  J. Commun. Soc.}, vol.~2, pp. 1431--1444, 2021.

\bibitem{9328518}
Z.~Ding, ``Harvesting devices' heterogeneous energy profiles and {QoS}
  requirements in {IoT}: {WPT-NOMA vs BAC-NOMA},'' \emph{IEEE Trans. Commun.},
  vol.~69, no.~5, pp. 2837--2850, 2021.

\bibitem{8761990}
Q.~Zhang, L.~Zhang, Y.-C. Liang, and P.~Y. Kam, ``Backscatter-{NOMA}: An
  integrated system of cellular and internet-of-things networks,'' in
  \emph{Proc. IEEE ICC}, 2019, pp. 1--6.

\bibitem{8807353}
S.~Zhou, W.~Xu, K.~Wang, C.~Pan, M.-S. Alouini, and A.~Nallanathan, ``Ergodic
  rate analysis of cooperative ambient backscatter communication,'' \emph{IEEE
  Wireless Commun. Lett.}, vol.~8, no.~6, pp. 1679--1682, 2019.

\bibitem{9220812}
Z.~Liu, G.~Lu, Y.~Ye, and X.~Chu, ``System outage probability of {PS-SWIPT}
  enabled two-way {AF} relaying with hardware impairments,'' \emph{IEEE Trans.
  Veh. Technol.}, vol.~69, no.~11, pp. 13\,532--13\,545, 2020.

\bibitem{9319204}
X.~Li, M.~Zhao, M.~Zeng, S.~Mumtaz, V.~G. Menon, Z.~Ding, and O.~A. Dobre,
  ``Hardware impaired ambient backscatter {NOMA} systems: Reliability and
  security,'' \emph{IEEE Trans. Commun.}, vol.~69, no.~4, pp. 2723--2736, 2021.

\bibitem{b1}
I.~S. {Gradshteyn} and I.~M. {Ryzhik}, \emph{Table of integrals, series, and
  products}, 7th ed. New York, NY, USA: Academic press, 2007.

\end{thebibliography}

\end{document}